\documentclass[epj]{webofc}
\usepackage[varg]{txfonts}  
\usepackage{slashed}
\usepackage{dsfont}
\usepackage{bm}
\def\beq{\begin{equation}}
\def\eeq{\end{equation}}
\def\bea{\begin{eqnarray}}
\def\eea{\end{eqnarray}}
\def\nn{\nonumber}

\def\q{{\bf q}}

\def\P{{\bf P}}
\def\Tr{\rm{Tr}}

\wocname{EPJ Web of Conferences}
\woctitle{CONF12}
\begin{document}
\selectlanguage{english}
\title{Charmed mesons at finite temperature and chemical potential}
%
%

\author{Fernando E. Serna\inst{1}\fnsep\thanks{\email{fernandoserna@ift.unesp.br}} \and
         Gast\~{a}o Krein\inst{1}\fnsep\thanks{\email{gkrein@ift.unesp.br}}
}


\institute{Instituto de F\'{\i}sica Te\'{o}rica, Universidade Estadual
Paulista \\
Rua Dr. Bento Teobaldo Ferraz, 271 - Bloco II, 01140-070 S\~{a}o Paulo, SP, Brazil
}
\abstract{%
We compute the masses of the pseudoscalar mesons $\pi^+$, $K^0$ 
and $D^+$ at finite temperature and baryon chemical potential.
The computations are based on a symmetry-preserving Dyson-Schwinger
equation treatment of a vector-vector four quark contact interaction.
The results found for the temperature dependence of the meson masses 
are in qualitative agreement with lattice QCD data and QCD sum
rules calculations. The chemical potential dependence of the masses
provide a novel prediction of the present computation. 
}
\maketitle
\section{Introduction}
\label{intro}
Heavy-light mesons, like the $D$ and $B$, are interesting QCD bound states:
they play an important role in the threshold dynamics of many of the so-called 
X,Y,Z exotic hadrons~\cite{Briceno:2015rlt} and also serve as a laboratory for 
studying chiral properties of the light $u$ and $d$ quarks in a medium at finite 
temperature~$T$ and baryon density~$\mu$ in different contexts. Great progress 
has been achieved in recent years in the study of their properties with lattice QCD 
methods, both in vacuum and at finite temperature~\cite{Skull}. In the continuum, 
however, although in principle the complex of the Schwinger-Dyson (DS) 
and Bethe-Salpeter (BS) equations provides an adequate framework, tremendous 
challenges still remain in describing simultaneously light- and heavy-flavored 
mesons in vacuum within a single model interaction--truncation 
scheme~\cite{{MT},{Krass},{Rojas:2014aka}}. On the other hand, there is pressing
need for different pieces of information on properties of such mesons both in
vacuum and at finite $T$ and $\mu$ for guiding experimental proposals at existing 
and forthcoming facilities. Examples are charmed-hadron production via $\bar{p}p$ 
annihilation processes~\cite{{Haidenbauer:2009ad},{Haidenbauer:2014rva},
{Haidenbauer:2015vra}}, $J/\Psi$-nuclear bound states~\cite{{Krein:2010vp},{Tsushima:2011kh}},
production in heavy-ion collisions of exotic molecules like the $\Delta\bar{D}^*$~\cite{Carames:2016qhr}.
Having this in mind, in the present contribution we extend to finite $T$ and $\mu$ a model that 
gives a good description of the spectrum and leptonic decay constants of pseudoscalar mesons in 
vacuum~\cite{F.E.S} . The model is based on a confining, symmetry-preserving treatment 
of a vector-vector four fermion contact interaction as a representation of the gluon's two-point 
Schwinger function used in kernels of DS equations, originally tuned to study the 
pion~\cite{GutierrezGuerrero:2010md}.

A potential problem with contact-interaction models is their nonrenormalizability, in that 
it can introduce gross violations of global and local symmetries because of ambiguities 
related to momentum shifts in divergent integrals. Here we use a subtraction 
scheme~\cite{{Battistel:1998tj}} that allows us to separate symmetry-violating 
parts in Bethe-Salpeter amplitudes in a way independent of choices of momentum routing in divergent 
integrals. The scheme has been used in the past within the Nambu--Jona-Lasinio (NJL) model in 
vacuum~\cite{Battistel:2008fd}, at finite $T$ and 
$\mu$ ~\cite{Farias} and, more 
recently~\cite{Farias:2016let}, it was used to explain the reason for the failure of the model 
to explain lattice results for the chiral transition temperature in the presence of a chiral 
imbalance in quark matter. 

We extend the subtraction scheme of Ref.~\cite{F.E.S} to finite $T$ and
$\mu$; the situation is more complicated than in the vacuum because Lorentz covariance
is broken at finite $T$ and $\mu$ and special care must be exercised to separate purely
divergent contributions from thermal effects, which are finite and do not need regularization. 
After setting up the scheme, we calculate the masses of the pseudoscalar 
mesons $\pi^+$, $K^0$ and $D^+$ at finite $T$ and $\mu$ and compare results with those
obtained recently using QCD sum rules~\cite{Suzuki:2015est} and those obtained earlier 
with the NJL model~\cite{{Blaschke:2011yv},{Gottfried:1992bz}}.

\section{Dyson-Schwinger and Bethe-Salpeter equations at finite $\boldsymbol{T}$ 
and $\boldsymbol{\mu}$ }
\label{sec-1}

The Dyson-Schwinger equation (DSE) for the full quark propagator $S_f(k)$ of flavor $f$ is given by 
(in Euclidean space)
\beq
S^{-1}_f(k) = i \slashed k + m_f 
+ \int\frac{d^4q}{(2\pi)^4} \, g^2 
D_{\mu\nu}(k-q) 
 \, \frac{\lambda^a}{2}\gamma_\mu 
S_f(q) \frac{\lambda^a}{2}\Gamma^f_\nu(q,k)~,
\label{eq:DSEqp}
\eeq
where $m_f$ is the current-quark mass, $D^{\mu\nu}$ the full gluon propagator, 
and $\Gamma^f_\nu$ the full quark-gluon vertex. The mass $m_{\rm PS}$ of a 
pseudoscalar (PS) meson with one light ($f=l$) quark and one heavy ($f=h$) quark is the eigenvalue 
$P^2 =  - m^2_{\rm PS}$ that solves the homogeneous Bethe-Salpeter equation (BSE) 
\beq
\Gamma^{lh}_{\rm PS}(k;P) = \int\frac{d^4q}{(2\pi)^4}K(k,q;P)
S_l(q_+)\Gamma^{lh}_{\rm PS}(q;P)S_h(q_-)~, 
\label{eq:BSEps}
\eeq
with $K(k,q;P)$ being the fully amputated quark-antiquark scattering kernel, 
where $q_\pm = q  \pm \eta_\pm P$ and $\eta_+ + \eta_- = 1$. 
At finite $T$ and $\mu$, the four dimensional momentum integrals in Eqs.~(\ref{eq:DSEqp}) and (\ref{eq:BSEps}) 
become 
\beq
\int\frac{d^4q}{(2\pi)^4} \, F(q)
\rightarrow \frac{1}{\beta}\sum^{\infty}_{m=-\infty}\int\frac{d^3q}{(2\pi)^3}\,F(\omega_m,\q)~,
\eeq
where $\omega_m=(2m+1)\,\pi/\beta$ are the fermionic Matsubara frequencies with $\beta =1/T$ and
$m\in \mathds{Z}$. 
The contact-interaction limit of full QCD is obtained by making the following replacements in 
Eqs.~\eqref{eq:DSEqp} and (\ref{eq:BSEps})
\beq
g^{2} D_{\mu\nu}(p-q) \rightarrow G \, \delta_{\mu\nu}, \hspace{0.7cm}
\Gamma^a_\mu \rightarrow \frac{\lambda^a}{2}\gamma_\mu, 
\hspace{0.7cm}
K(k,q;P) = - G \, \left(\frac{\lambda^a}{2}\gamma^\mu\right) \otimes
\left(\frac{\lambda^a}{2}\gamma_\mu\right)~,  
\label{contact}
\eeq
where $G$ is an effective coupling constant with dimensions of (length)$^2$. In this limit,
Eq.~(\ref{eq:DSEqp}) becomes
\beq
S^{-1}_f({\q}, \omega_m) = i {\bm\gamma}\cdot \q + i \gamma_4\,\omega_m + M_f~,
\eeq
with $M_f = M_f(T,\mu)$ given by (the gap equation)
\beq
M_f = m_f + \frac{16}{3} G \, \left\{I_{\rm quad}(M_f)  - \int\frac{d^3 q}{(2\pi)^3} \frac{M_f}{2E_f(\q)}
\left[f^-(E_f(\q)) + f^+(E_f(\q))\right]\right\}~,
\label{Tgap}
\eeq
where 
\beq
I_{\rm quad}(M^2) = \int\frac{d^4 q}{(2\pi)^4} \,\frac{1}{q^2 + M^2}~,
\hspace{0.5cm}f^\pm(E) = \frac{1}{1 + e^{\,\beta E^\pm}}~,
\label{Iquad}
\eeq
with $E^{\pm}_f(\q) = E_f(\q) \pm \mu$, and $E_f(q) = (\q^2+M^2_f)^{1/2}$. 
The Bethe-Salpeter amplitude (BSA) contains only pseudo-scalar and pseudo-vector 
components:
\beq
\label{contactBSA}
\Gamma^{lh}_{\rm PS}(P)=\gamma_5\left[iE^{lh}_{\rm PS}(P)+\frac{1}{2M_{lh}}\gamma\cdotp 
P\,F^{lh}_{\rm PS}(P)\right]~,
\eeq
with $P = (\nu_n, \P)$, $\nu_n = 2 n\pi/\beta$, and the factor $M_{lh}=M_lM_h/(M_l+M_h)$, 
where $M_l$ and $M_h$ are solutions of Eq.~(\ref{Tgap}), is introduced for convenience. Using 
this in Eq.~(\ref{eq:BSEps}), the BSE of 
can be written~\cite{F.E.S} as a matrix equation involving the amplitudes $E^{lh}_{\rm PS}$ and $F^{lh}_{\rm PS}$. 
For comparison with earlier results in the literature that use the Random Phase Approximation (RPA), one needs to 
use only the pseudoscalar component $E^{lh}_{\rm PS}$ in $\Gamma^{lh}_{\rm PS}(P)$. The meson mass
is obtained taking $\omega = m_{\rm PS}$ in the expression 
\beq
0 = 1 - \frac{G}{3} \,\Pi_{\rm PS}(\nu_n=-i\omega,\P=0)~,
\eeq 
with
\beq
\label{PI_T}
\Pi^{lh}_{\rm PS}(\nu_n,\P) = -\frac{1}{\beta}\sum^{\infty}_{m=-\infty}
\int\frac{d^3 q}{(2\pi)^3}{\Tr}\left[\gamma_5 \gamma_{\mu} S_l(\q_+,\omega_+)
\gamma_5\,S_h(\q_-,\omega_-)\gamma_{\mu}\right]~,
\eeq
where $\q_{\pm} = \q \pm \eta_{\pm} \, \P$ and  $\omega_{\pm} = \omega_m \pm \chi_{\pm} \, \nu_n$,
where $\chi_{\pm}\in \mathds{Z}$. After taking a Dirac trace we can rewrite Eq.~\eqref{PI_T} as a sum 
of two terms: one that contains ultraviolet divergences
\beq
\Pi^{lh}_{\rm PS}(P)|_{\rm div} = 8 \int \frac{d^4q}{(2\pi)^4} \Bigg\{\frac{1}{q^2_+ + M^2_l}+\frac{1}{q^2_- + M^2_h}
-\left[ P^2 + \left(M_h - M_l\right)^2 \right]\frac{1}{(q^2_+ + M^2_l)(q^2_+ + M^2_h)}\Bigg\}~,
\label{Pi-div}
\eeq
and another that is finite, given in terms of the Fermi-Dirac distributions
$f^\pm(E)$ defined in Eq.~(\ref{Iquad}):
\bea
\Pi^{lh}_{\rm PS}(P)|_{\rm fin}&=&  - 4 \int\frac{d^3q}{(2\pi)^3}
\Bigg\{ \frac{f^+(E_l(\q_+))+f^-(E_l(\q_+))}{E_l(\q_+))}
+\frac{f^+(E_h(\q_-)) + f^-(E_h(\q_-))}{E_h(\q_-)}
\nn\\[0.3true cm]
&+&\frac{P^2 + (M_l-M_h)^2}{2E_l(\q_+)E_h(\q_-)}
\Bigg[\frac{2E_h(\q_-)f^-(E_l(\q_+))}{ [\omega - E_l(\q_+)]^2- E^2_h(\q_-)}
+\frac{2E_l(\q_+)f^-(E_h(\q_-))}{ [\omega + E_h(\q_-)]^2- E^2_l(\q_+)}
\nn\\[0.3true cm]
&+&\frac{2E_h(\q_-)f^+(E_l(\q_+))}{ [\omega + E_l(\q_+)]^2- E^2_h(\q_-)}
+\frac{2E_l(\q_+)f^+(E_h(\q_-))}{ [\omega - E_h(\q_-)]^2- E^2_l(\q_+)}
\Bigg]\Bigg\}~.
\label{Pi-fin}
\eea
In Eqs.~(\ref{Pi-div}) and (\ref{Pi-fin}), $P = (-i\omega,\P)$. The problem with symmetry violations alluded to in the 
Introduction comes from the divergent part: it depends on the choice made
for the partition of the momenta in the loop integral when using a cutoff regularization, 
i.e. it is not independent of $\eta_\pm$. Our scheme~\cite{F.E.S} to obtain symmetry-preserving 
expressions is to perform subtractions in divergent integrals: 
\bea
\frac{1}{q^2_{\pm} + M^2_{l,h}} &=& \frac{1}{q^2 + M^2}
+\left(\frac{1}{q^2_{\pm} + M^2_{l,h}}-\frac{1}{q^2 + M^2}\right) 
=  \frac{1}{q^2 + M^2} - \frac{(q^2_\pm - q^2 + M^2_{l,h} -M^2 )}
{\left(q^2 + M^2\right)(q^2_\pm + M^2_{l,h})},
\label{subtr-1}
\eea   
with $M$ being an arbitrary subtraction mass scale. One performs as many subtractions as necessary
to obtain one finite integral. The final result is~\cite{F.E.S}
\bea
\Pi^{lh}_{\rm PS}(P)|_{\rm div} &=&\frac{3}{G}
-\frac{3}{2G}\left(\frac{m_l}{M_l}+\frac{m_h}{M_h}\right)
- 8 \left[P^2  + (M_l-M_h)^2 \right]
\left[ I_{\rm log}(M) - Z_0(M^2_l,M^2_h,P^2,M^2) \right]
\nn\\[0.1true cm]
&&\, + 8 \, (\eta^2_+ + \eta^2_-) \, A_{\mu\nu}(M^2) \, P_\mu P_\nu~,
\eea 
where $I_{\rm log}(M^2)$ is a divergent integral and $Z_0(M^2_l,M^2_h,P^2,M^2)$ 
is a finite integral
\beq
I_{\rm log}(M^2) = \int\frac{d^4 q}{(2\pi)^4} \,\frac{1}{\left(q^2 + M^2\right)^2}~,
\hspace{0.5cm}
Z_0(M^2_l,M^2_h,P^2,M^2)=\frac{1}{(4\pi)^2}\int^1_0dz
\ln\left[\frac{H(z,P^2)}{M^2}\right]~,
\label{Ilog}
\eeq
with $H=z(z-1) P^2  - z (M^2_l-M^2_h) + M^2_l$, and $A_{\mu\nu}(M^2)$ is another
divergent integral
\beq
A_{\mu\nu}(M^2) = \int \frac{d^4 q}{(2\pi)^4}  
\frac{ 4q_\mu q_\nu - (q^2 + M^2) \delta_{\mu\nu} }{ (q^2 + M^2)^3 }~.
\label{A_munu}
\eeq
The finite integral in Eq.~(\ref{Ilog}) is obtained by integrating over momentum, 
removing any regularization implicitly assumed. Clearly, the term proportional to 
$A_{\mu\nu}(M^2)$ is not independent of $\eta_\pm$ and, therefore, violates translation 
symmetry. Similar expressions appear also in Ward-Takahashi identities~\cite{F.E.S}. 
Since there are regularizations schemes, like dimensional regularization, where 
$A_{\mu\nu}(M^2) = 0$ automatically, it is a natural
prescription for obtaining symmetry-preserving amplitudes to demand the vanishing of 
$A_{\mu\nu}(M^2)$ (and of all other similar terms that appear in other amplitudes, 
see e.g. Ref.~\cite{F.E.S}), independently of the regularization used to regulate 
$I_{\rm quad}(M^2)$ and $I_{\rm log}(M^2)$. 

\begin{figure}[t]
\begin{center}
\includegraphics[scale=0.6]{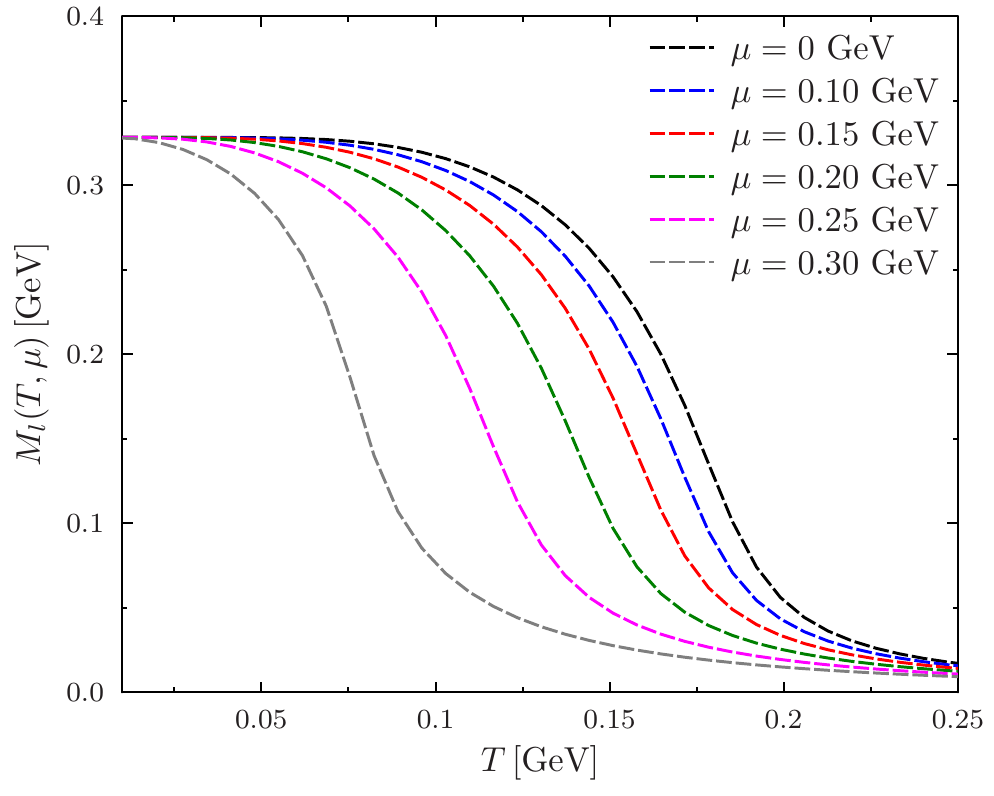}\hspace{1.0cm}
\includegraphics[scale=0.6]{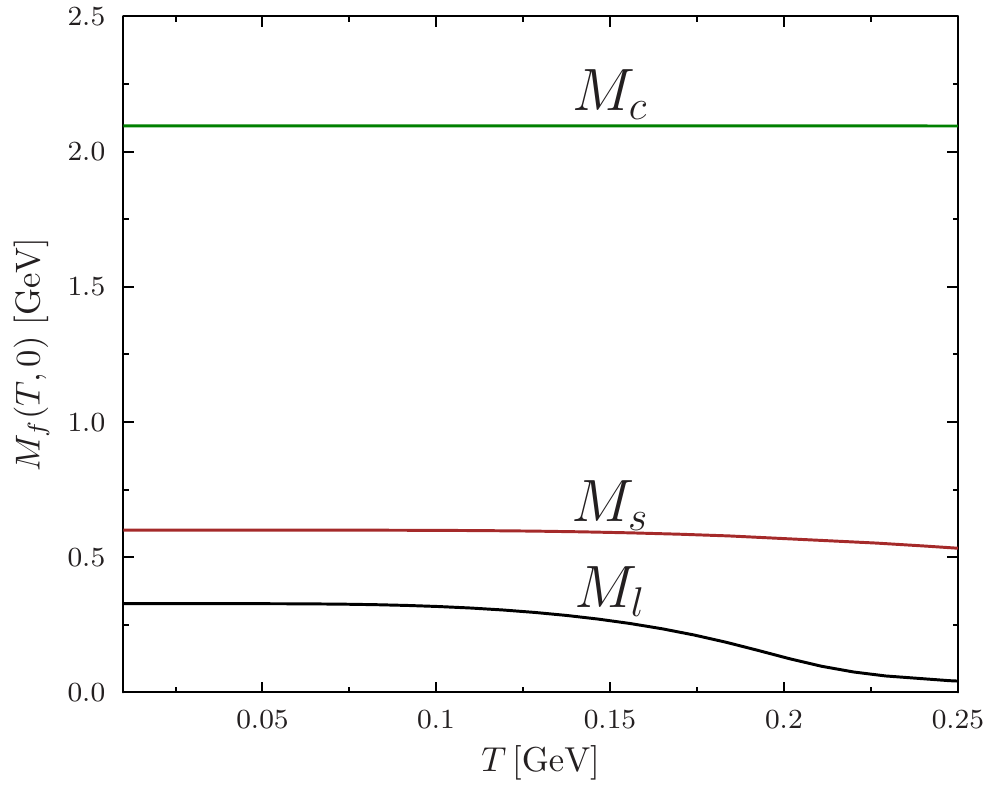}
\end{center}
\vspace{-0.5cm}
\caption{Constituent quark mass as function of $T$ for different values of $\mu$.}
\label{figI}
\end{figure}

The mass scale $M$ in Eq.~(\ref{subtr-1}) is arbitrary;
it appears in the divergent integrals $I_{\rm log}(M^2)$ and $I_{\rm quad}(M^2)$ and 
in the finite integral $Z_0(M^2_l,M^2_h,P^2,M^2)$. Since the model is nonrenormalizable, 
the integrals $I_{\rm log}(M^2)$ and $I_{\rm quad}(M^2)$ cannot be removed, of course.
One can use an explicit regulator to evaluate the integrals and fit the regulator to 
physical quantities, like the quark condensate or hadron masses, or one could also fit the 
integrals directly to physical quantities. In either case, the fit to physical quantities 
is $M$-dependent, that is, physical quantities would ``run with $M$'', very much like in
renormalizable quantum field theories where all masses and other quantities are running
functions of a mass scale that enters the theory via the regularization scheme. Here we
present results using an explicit three-dimensional cutoff $\Lambda$ to regulate the 
divergent integrals $I_{\rm log}(M^2)$ and $I_{\rm quad}(M^2)$, and take $M=M_h$, for 
simplicity{\textemdash}further discussions on this will be presented elsewhere.

\section{Numerical results and conclusions}

The free parameters are: $G$, $\Lambda$ and $m_f$. Taking $\Lambda = 0.653~{\rm GeV}$, 
$G\Lambda^2 = 19.26$, $m_l=m_u = m_d = 0.005$~GeV, $m_s = 0.161$~GeV and $m_c = 1.544$~GeV,
one obtains for the meson masses in vacuum: $m_\pi = 0.139$~GeV, $m_{K^0} = 0.493$~GeV 
and $m_{D^+} = 1.864$~GeV. It is worth mentioning that we obtain for the constituent quark
masses in vacuum the following values: $M_l=0.328$~GeV, $M_s = 0.599$~GeV and $M_c=2.095$~GeV.
The results for the $T$ dependence of $M_l$ for different values of $\mu$, obtained from
the DSE in Eq.~(\ref{Tgap}), are shown in the left panel of Fig.~\ref{figI}. Clearly, as
$\mu$ increases, the (pseudo) critical temperature for chiral restoration decreases
as $\mu$ increases, as expected. On the right panel of the figure, the temperature dependence
of $M_l$, $M_s$ and $M_c$ for zero chemical potential: also as expected, as the current
quark mass increases, the effect of the temperature becomes less important, even close to 
and above the pseudocritical temperature.

\begin{figure}[t]
\begin{center}
\includegraphics[scale=0.6]{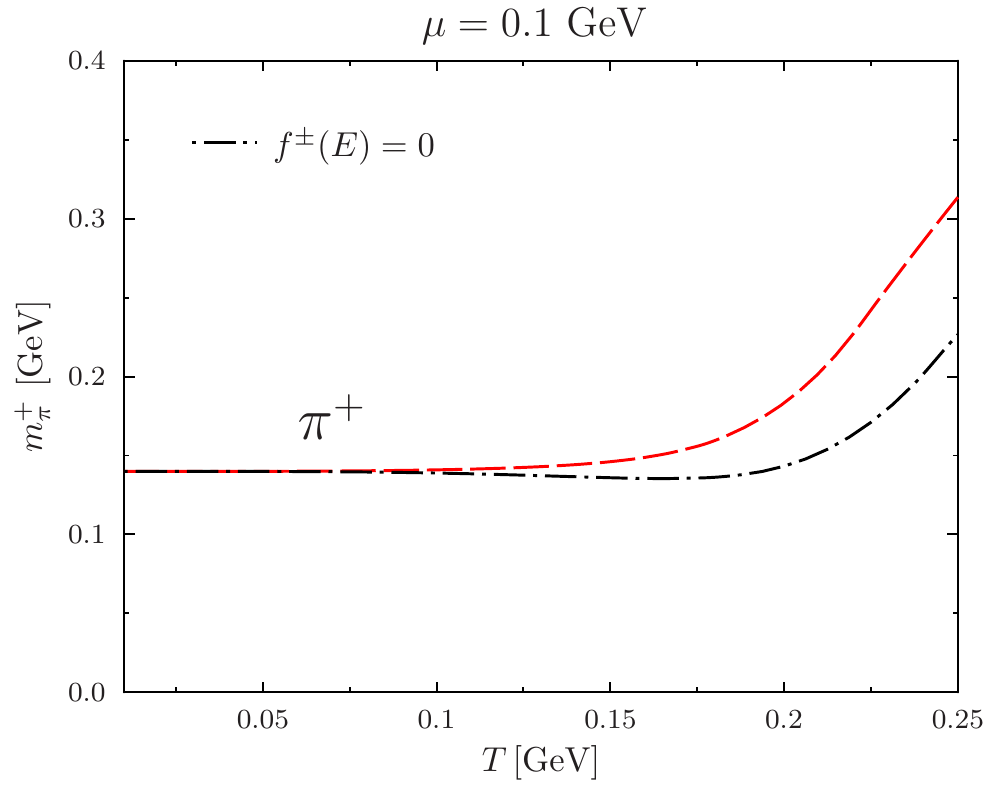}\hspace{0.25cm}
\includegraphics[scale=0.6]{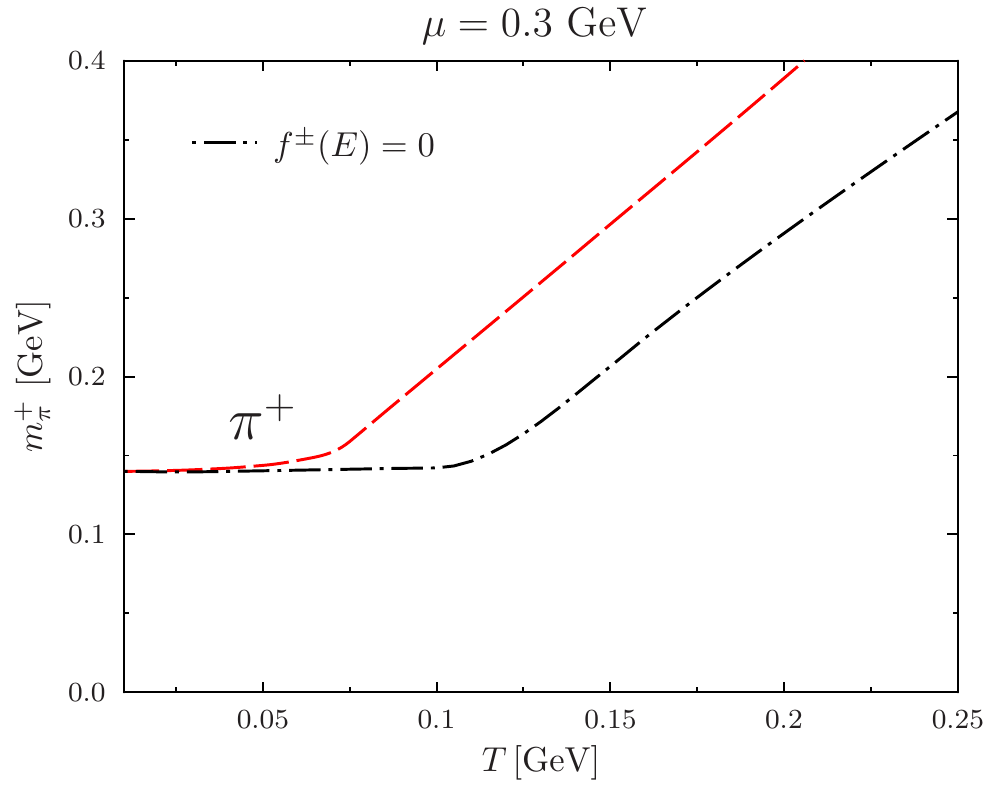}\vspace{0.25cm}
\includegraphics[scale=0.6]{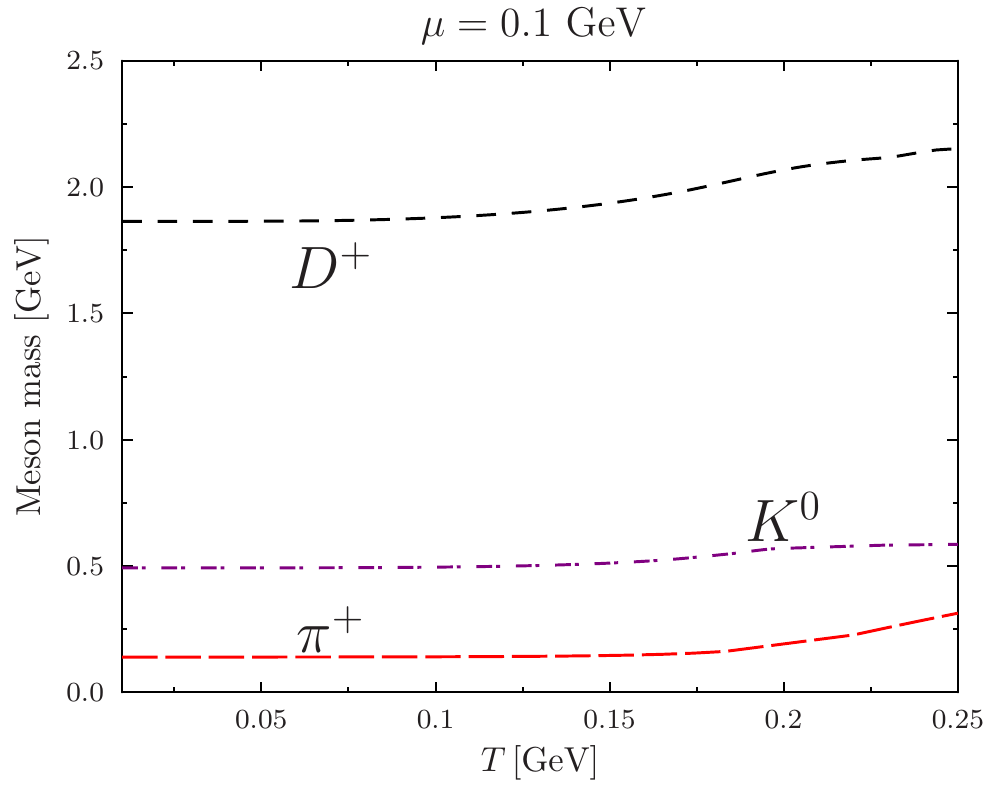}\hspace{0.25cm}
\includegraphics[scale=0.6]{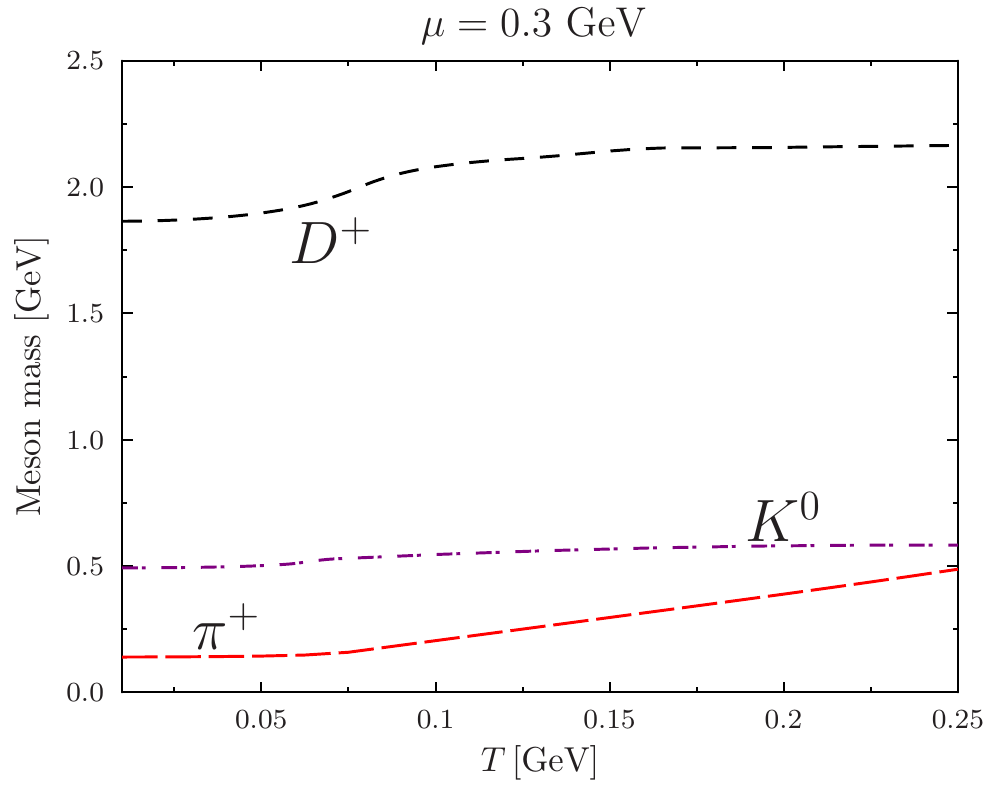}
\end{center}
\vspace{-0.5cm}
\caption{Meson masses as functions of $T$ for two typical values of $\mu$. Top panels:
$m_{\pi^+}$ calculated with and without the Fermi-Dirac distributions $f^\pm(E)$ in 
the BSE.
Bottom panels: masses calculated without the $f^\pm(E)$ in the BSE.}
\label{figII}
\end{figure}

The last point raises the question on the importance of the Fermi-Dirac distribution
functions, $f^\pm(E)$, in the BSE for the masses of the mesons{\textemdash}physically,
they represent the contributions of thermally activated quark-antiquark pairs in the 
bound state. The answer to this question is shown in the top panels of Fig.~\ref{figII}:
for sufficiently low values of $\mu$, the $f^\pm(E)$ play no significant role for $T$ 
smaller than the pseudocritical temperature and can, to a good approximation, be neglected. 
This is an important feature, as it simplifies considerably the calculations of thermal 
effects on hadron masses, as all $T$ and $\mu$ effects below the crossover are captured 
by the $T$ and $\mu$ dependence of the constituent quark masses. In the bottom panel of 
Fig.~\ref{figII}, we show the results for the mesons masses neglecting the Fermi-Dirac 
distributions: the results are in good quantitative agreement with early 
calculations using the NJL model~\cite{{Blaschke:2011yv},{Gottfried:1992bz}}. It also 
agrees with a very recent calculation using a chiral constituent quark model, using
as input $T$ and $\mu$ dependent quark masses and quark-meson couplings~\cite{Carames:2016qhr},
Finally, our results are also in qualitative agreement with a recent calculation using QCD 
sum rules~\cite{Suzuki:2015est}{\textemdash}this last reference points to earlier calculations,
that obtained the opposite trend or the $D$ meson mass. We also mention that recent lattice
results~\cite{Skull} show that the masses of $D$ mesons increase, and become broader.

As perspectives, concrete calculations of production rates of exotic hadronic molecules 
in heavy-ion collisions~\cite{{Briceno:2015rlt},{Carames:2016qhr}} and of transport properties
of charmed hadrons~\cite{Ghosh:2015mda} are top priority. In addition, it would be important 
to contrast results using confining chiral models. Of particular 
interest to us are those models inspired in QCD formulated in Coulomb 
gauge~\cite{{Bicudo:1991kz},{Fontoura:2012mz}}, and those based on chiral soliton 
models~\cite{{Krein:1988sb},{Krein:1988vh}}.

\begin{acknowledgement}
We thank B. El-Bennich for valuable discussions.
Work partially by Conselho Nacional de Desenvolvimento 
Cient\'{\i}fico e Tecnol\'ogico - CNPq, Grants No. 305894/2009-9 (G.K.)
and 140041/2014-1 (F.E.S.) and Funda\c{c}\~ao de Amparo \`a Pesquisa do Estado 
de S\~ao Paulo-FAPESP, Grant No. 2013/01907-0 (G.K.).  
\end{acknowledgement}


\end{document}